# The effect of monetary incentives on sociality induced cooperation

Tatiana Kozitsina (Babkina)[1,2*], Alexander Chaban[1], Evgeniya Lukinova[3], Mikhail Myagkov[4,5]

[1]Moscow Institute of Physics and Technology (State University), 9 Institutskiy per., Dolgoprudny, Moscow Region, 141701, Russian Federation

[2]Federal Research Center Computer Science and Control, Russian Academy of Sciences, Moscow, 119333, Russian Federation

[3]NYU-ECNU Institute of Brain and Cognitive Science at New York University Shanghai, 1555 Century Ave, Pudong, Shanghai, China

[4]Tomsk State University, 36 Lenina Avenue, Tomsk, 634050, Russian Federation

[5]University of Oregon, 1585 E. 13th Avenue, Eugene, Oregon, 97403, United States

* Corresponding author

E-mail: babkinats@yandex.ru

*Abstract*— **This paper examines how the group membership fee influences the formation of groups and the cooperation rate within the socialized groups. We found that monetary transactions do not ruin the establishment of social ties and the formation of group relations.**

## I. INTRODUCTION

The development of all humanity is due to the interaction between people. Human interaction, in turn, is tightly linked to the transactions that happen between people, which are rarely accomplished with pure altruistic motives. The life of the modern society is driven by the rule that everything has a price and must be paid for. However, is it really so? According to Simmel [1], economic exchange has the same nature as social interaction. So, is there a difference in social behavior, if comparing a situation without money to a situation with money?

Often humans are considered to be inherently asocial, rational, and selfish [2–4]. According to this view, a person always acts in a way to maximize his benefits: the larger the cost, the more selfish the behavior is. On the other hand, according to Fiske [5,6], there are four forms of human relationships, and not all of them are based on mutual benefit. Thus, money and prosocial behavior can exist together, for example, this can be directly observed in particular societies [7].

There is evidence [8–12], that the cooperation rate in a laboratory group where strangers can organize in socialized groups is comparable to the level of cooperation in groups that have long-standing ties. Authors postulate that sociality, even in a very minimal form, serves as a natural mechanism to promote sustainable cooperation among group members [9]. These findings directly relate to the recent results indicating that the visual contact before the experiment facilitates the cooperation [13]. Moreover, the brain area associated with self-control can modulate the valuation system for in-group cooperation, thus, forcing social behaviors in the newly formed groups [14].

However, little is known about how the group membership fee influences the formation of the groups and the cooperation rate within the newly socialized groups. So, in this paper we report on a series of experiments where the membership fee is required to enter the group; if you do not want to pay, you stay out. We compare it to another set of experiments where the monetary fee is absent. Taken together, this will sort out 'the membership fee problem', or how to maintain the collective action given the monetary confounds, which are the very essence of the human relationships today.

The scholarship has not reached consensus on this matter. For example, the study [15] demonstrates that segregation is arranged in such a way that the person seeks the group, in which he expects the biggest benefit. On the other hand, researchers [16] claim that any movement between the groups adversely affects the cooperation. It turns out that the highest cooperation is expected in the group set up by personal preference with a strictly fixed number of participants. The paper [17] states that the participants tend to overpay in the auctions. Similarly, in [18] participants played the Prisoner's Dilemma game (PD) against strangers, then socialized within a group and repeated the PD in socialized groups, after that, they were proposed to play the PD game with auctions every 5 rounds. To stay in-group, participants were obligated to pay the fee. The authors conclude [18] that in-group participants maintained the cooperation level high, but out-group cooperation rate declined to the level observed in the game against strangers. Interestingly, people tend to join the group with cooperators, even without directly observing others behavior [19]. So, if participants can recognize the cooperative intentions in the group, they will join the group. According to the previous findings, most people change their strategies in the social group from defective to cooperative. Thus, most people expect the cooperation in the group. Will this work in this way in fact? Even in the losses framework, people desire to be inside the social groups [20]. Based on this, we can also suppose that if people are afraid of defectors, they will, all the same, want to join the group.

Summing it up, we raise the following research questions: how a payment requirement to join a group will influence the social behavior?

To answer our questions, we created the experimental sessions that we carefully explain in Materials and Methods section. Our main purpose was to understand the influence of a monetary fee on the socialization in experiments, where group membership fee is involved or is absent. With this study we both explore the influence of money on the already socialized groups and on the newly formed groups.

## II. MATERIALS AND METHOD

### A. Participants

Subjects (16 experiments, N = 192, 130 males) for the experiment were recruited from the students at the Moscow Institute of Physics and Technology (MIPT). The MIPT Experimental Economics laboratory was used to carry out all experiments. Each experiment consisted of 12 students, pre-selected before the experiment to be unfamiliar with one another. A specialized tool to design and carry out group experiments in experimental economics, z-Tree developed at the University of Zurich, was used [21]. After the end of each treatment, participants provided feedback about the experiments, received payments and left the experimental facility. The study procedures involving human participants were approved by Tomsk State University Human Subjects Committee. Written informed consents were obtained from participants. Experimental data are readily available on Harvard Dataverse:
https://dataverse.harvard.edu/privateurl.xhtml?token=2b783457-1365-4026-900d-586c45d681b7

### B. Procedure

Twelve participants were invited for each experiment. All participants received written and verbal instructions for the experiment. Experimenters announced that all points that the participants will win during the experiment will be converted to the real money. Each experiment was divided into the following three phases. In the first and the third phases, participants played the iterated Prisoner's Dilemma game (PD) [22] and/or the iterated Investment Game (IG) [23].

In PD there are two strategies: Cooperation (Up or Left) or Defection (Down or Right) (Table 1). If two players both choose to either cooperate or defect, they gain the same points, 5, for Cooperation and a smaller gain, 1, for Defection, respectively. If one of the players cooperates and another defects, the cooperator gains a smaller reward, 0, but the defector takes a larger reward, 10. According to the expected value of the reward, Defection is more profitable than Cooperation in any partner's choice, but mutual Cooperation is more profitable for both than mutual Defection [24]. Cooperation rate in PD is calculated individually as the proportion of cooperative choices in all games a player takes part. According to the rules during the game a player can be either a row-chooser or a column-chooser, i.e. choosing between rows (Up and Down) or columns (Left or Right).

TABLE I.    PRISONER'S DILEMMA PAYOFFS

| Payoffs | Cooperation | Defection |
|---|---|---|
| Cooperation | 5, 5 | 0, 10 |
| Defection | 10, 0 | 1, 1 |

In IG the first subject (trusting) sends (trusts) to the opponent any number of points from the total of 10. The opponent (the second subject or grateful) gets this sum multiplied by three. After that, the opponent can return any part of the received amount (reciprocity) to the first subject. Thus, during the game participant can play either of two roles: trusting (as the first subject in the example or grateful (as the second subject in the example). In the totally mixed Nash equilibrium, it is not rational to return something, and therefore it is not rational to trust, which leads to a zero result for both participants. Average trust for some subject in IG is defined as the average amount of points that the subject sends to the opponent. Similarly, reciprocity is the average amount of points that the subject returns to the opponent.

There were three studies to compare socialization without money treatment (later simple socialization) with socialization with money treatment: Study 1 (8 experiments, N = 96, 59 males) used both PD, IG, and socialization with money treatment; Study 2 (4 experiments, N = 48, 35 males) used PD only and simple socialization, and Study 3 (4 experiments, N = 48, 36 males) used IG only and simple socialization.

### C. Experimental design

Each experiment consisted of three parts.

Anonymous Game

Study 1

Participants played the two-person PD with a random human partner to make the game anonymous. Participants were randomly paired with an anonymous partner each round of the game and randomly alternated roles between column chooser and row chooser for the PD. Then the participants played the IG with a random human partner. Participants were randomly paired with an anonymous partner each round of the game and randomly alternated roles between the trusting and grateful subjects. This game phase lasted for 22 rounds for both games.

Study 2

Participants played the two-person PD with a random human partner. Participants were randomly paired with an anonymous partner each round of the game and alternated roles on

This research was supported by The Tomsk State University competitiveness improvement program and by the grant RFBR 19-01-00296A subsequent trials between column chooser and row chooser for the PD. This game phase lasted for 22 rounds.

Study 3

Participants played the two-person IG with a random human partner. Participants were randomly paired with an anonymous partner each round of the game and alternated roles on subsequent trials between trust and reciprocity roles for the IG. This game phase lasted for 22 rounds.

Socialization phase

The procedure from socialization phase was the similar to the procedures used in previous work [8].

Socialization with Money treatment phase for Study 1

All 12 subjects participated in the ice-breaker [8]: first of them told his/her name and an adjective that started from the same letter, second subject repeated first subject's name and adjective, and said his/her own name, and adjective, and so on till the last participant said all names and adjectives in order. Then, in a reverse order each participant shared his/her life facts. After that, participants were asked who wanted to become a group leader. Then, information on the number of points earned through the Anonymous Game phase of each participant became available to all participants. Participants (except the group leaders) were asked to indicate how many points (from 0 to 50) they want to pay to be in a group with one or the other group leader. Then, the bids were divided between group leaders and placed in descending order (for every leader). For both groups (with leaders) separately only the three participants with the highest bids were able to join the group with the leader of their choice and made a payment (only once) that was equal to the size of their bid. This way two groups of four players (in-group) were organized. The rest four participants whose bids were not high enough to join the group of their choice formed a third group of four players (out-group) (Fig 1). In case, if there was no possibility to divide participants in accordance with the rules, we asked them (or part of them) to repeat the group-choice procedure without announcing the bids and the results of the previous group division. The participants get to choose a group, which easily satisfies the minimal group requirement and social identity theory [25,26]. After the groups were formed, the in-groups were allowed to socialize, communicate, and decide on the group name (Group Socialization), whereas members of the out-group were not allowed to talk with each other and were separated from each other and the in-groups.

Socialization phase for Study 2 and Study 3

All 12 subjects participated in the ice-breaker: first of them told his/her name and an adjective that started from the same letter, second subject repeated first subject's name and adjective, and said his/her own name, and adjective, and so on till the last participant said all names and adjectives in order. Then, in a reverse order each participant shared his/her life facts. After that, participants were asked who wanted to become a group leader. The rest of participants one-by-one revealed their group preference. As a result, participants were divided into two groups of six players. Here, the participants also get to choose a group, which easily satisfies the minimal group requirement and social identity theory [25,26].

Socialized Game phase

Study 1

During this phase, Socialized Game with a Fee participants played the PD and IG with a random human partner from their group of four. Their partner changed each round of the game. There was a total of 18 rounds of PD and 18 rounds of IG in this game phase.

Study 2

During this phase, Socialized Game participants played the PD with a random human partner from their group of six. Their partner changed each round of the game. There was a total of 18 rounds of PD in this game phase.

Study 3

During this phase, Socialized Game participants played the IG with a random human partner from their group of six. Their partner changed each round of the game. There was a total of 18 rounds of IG in this game phase.

The total profit is the sum of Anonymous Game phase plus Socialized Game phase, and minus payment in socialization with money treatment phase for Study 1 (approximately 1000 RUR).

According to the division process, the in-group is the group with group socialization; the out-group is the group without group socialization for Study 1.

III. RESULTS

*A. The requirement to pay the membership fee for joining the group does not annul the Socialization effect*

Firstly, let's consider the behavioral results from Studies 2 and 3 because these experiments contain simple socialization. The main question to answer using these studies: is there any difference in behavior before (anonymous game) and after (socialized game) socialization? From Table 2 the cooperation rate in PD increases from 0.22 to 0.64 in Study 2 (Wilcoxon signed rank sum test, p-value = 1.214e-14). The average level of trust in IG also increases from 4.36 to 8.26 in Study 3 (Wilcoxon signed rank sum test, p-value = 1.102e-08). So, simple socialization significantly enhances both cooperation and trust.

Secondly, Study 1 differs from Studies 2 and 3 because it contains the analogue of simple socialization but with money treatment. After socialization with money treatment two groups are socialized (in-group) and one group is refused a socialization (out-group). For PD game and IG there is a significant increase in the cooperation rate and average trust from before socialization to after socialization for in-groups (PD: Wilcoxon signed rank sum test, p-value < 2.2e-16; IG: p-value = 1.102e-15). However, for out-groups these values are not different between stages (PD: Wilcoxon signed rank sum test, p-value = 0.189; IG: p-value = 0.296). Moreover, there is a significant difference in cooperation rate and average trust between in-groups and out-groups (PD: Wilcoxon rank sum test, p-value = 1.102e-09; IG: p-value = 1.611e-06).

TABLE II. DESCRIPTIVE STATISTICS OF THE RESULTS FROM STUDIES 1, 2, AND 3

| Study | Game | Anonymous Game | | Socialized Game | |
|---|---|---|---|---|---|
| | | M | SD | M | SD |
| Study 1 | PD | 0.281 | 0.246 | In-group: 0.796 | 0.298 |
| | | | | Out-group: 0.321 | 0.236 |
| | IG | 5.102 | 3.116 | In-group: 9.226 | 2.168 |
| | | | | Out-group: 5.775 | 3.833 |
| Study 2 | PD | 0.219 | 0.220 | 0.639 | 0.367 |
| Study 3 | IG | 4.361 | 3.398 | 8.225 | 2.429 |

Thirdly, the comparison of the results from Study 1 with Studies 2 and 3 postulates the following: there are no significant differences across studies for Anonymous game (PD: Wilcoxon rank sum test, p-value = 0.071; IG: p-value = 0.204) (Table 2). The difference in cooperation rate after socialization between Study 2 and Study 1 (in-group) is not so big (0.16) but it is significant (Wilcoxon rank sum test, p-value = 0.004). Similarly, the difference in average trust after socialization between Study 3 and Study 1 (in-group) is tiny (it equals to 1) but it is significant (Wilcoxon rank sum test, p-value = 0.003). But the general trends in these results indicate that socialization with money treatment works in the similar way as the simple socialization.

To summarize, despite the fact that the in-group members pay the membership fee, this does not affect Group Socialization and their cooperation. The requirement to pay the membership fee for joining the group does not annul the Socialization effect. Although the out-group participants are in the better game conditions than those in-group, because they do not need to pay the fee, but without Socialization, they show just a low basic level of cooperation.

*B. There is a tendency to cooperate and trust more in case of a larger membership fee (Study 1)*

The correlation between the in-group cooperation and bids which participants paid to join the group is equal to 0.42 (p = 0.003, Spearman's rank correlation). The correlation between the in-group trust and bids which participants paid to join the group is equal to 0.3353 (p = 0.03, Spearman's rank correlation).

This indicates that monetary transactions do not make people more asocial and individual, but vice versa, higher group fees make them even more social.

*C. Dependence of trust on reciprocity is more pronounced in-group rather than out-group or during Anonymous game phase*

The results of IG also can be compared by the dependence of the amount the first subject sends to the opponent on the reciprocal sum the second subject sends back. During Anonymous game phase, this coefficient is equal to 0.972 (Table 3, Model_1). Despite the fact that sum of trust is tripled for the second subject; the reciprocity is smaller than trust. After Money Socialization inside the in-group the dependence increases to 1.55 (Table 3, Model_2). It indicates the influence of sociality on decision making. For comparison, this coefficient for the out-group is equal 1.19 (Table 3, Model_3).

TABLE III. LINEAR REGRESSION ANALYSIS ASSESSING THE RELATIONSHIP BETWEEN RECIPROCAL GRATITUDE (DEPENDENT VARIABLE) AND TRUST (INDEPENDENT VARIABLE) DURING ANONYMOUS GAME PHASE

| Reciprocal gratitude | Model_1 | Model_2 | Model_3 |
|---|---|---|---|
| Trust to opponent | 0.972*** | 1.550*** | 1.196*** |
| | (0.0476) | (0.147) | (0.109) |
| Constant | -1.240*** | -2.362* | -1.482* |
| | (0.284) | (1.394) | (0.749) |
| Observations | 89 | 56 | 28 |
| R-squared | 0.828 | 0.673 | 0.823 |

Notes: Model_1 – for Anonymous game phase, Model_2 – for Money Socialized phase In-group, Model_3 – for Money Socialized game phase Out-group. Standard errors in parentheses: *** $p<0.01$, ** $p<0.05$, * $p<0.1$.

This result also indicates the influence of socialization on decision-making: participants after socialization tend to choose the collective strategies as opposed to individualistic rational ones.

IV. CONCLUSION

We carried out experiments to study how people interact with each other with and without social context. We have found that the membership fee does not change the overall level of cooperation and trust. It agrees with other work [27], promoting the idea that the existence of the competitive free markets in a country facilitates trust between citizens. People tend to stay within their social groups and interact cooperatively there. It turns out that socialization helps to re-create such a social environment in the lab for individuals that do not know each other [28]. In the out-group people change their behavior to non-cooperative. Participants that were more selfish with unknown people or without social ties tend to be more social inside newly formed group (consistent with the minimal group paradigm) even if it costs some money. Moreover, it was previously found that movement between groups negatively influenced the cooperation [16], instead, our paper ensures that socialization even with money treatment can smooth this effect.

We found an interesting behavioral feature: the more subjects pay for the entry into the group, the more cooperation and trust is observed. So, the requirement to pay membership fee might increase social responsibility: people tend to be more socially attractive following some heuristics [29].

In addition, we found that the membership fee as a division mechanism of people into groups has no effect on the sociality, the cooperation and the trust level between members. However, the groups formed without a fee and without socialization show the low level of cooperation and trust. It indicates that socialization gives the highest boost to collective action regardless of the membership fee presence or absence. Studying this more might be helpful for the organization of online-closed communities with paid content to improve the value of the content to the participants, to upgrade the models of opinion formation, and to forecast the effects of new regulations introduction [30-34].

ACKNOWLEDGMENT

We thank Rinat Yaminov for writing the programing code for experiments, Ann Sedush for technical help in conducting experiments, Olga Menshikova and Ivan Menshikov for thoughtful comments, Anastasiia Nikolaeva-Aranovich for the manuscript correction. This research was supported by The Tomsk State University competitiveness improvement program and by the grant in Russian Foundation for Basic Research (the grant number is 19-01-00296A).